# ELECTRON CLOUD MEASUREMENTS IN FERMILAB BOOSTER*

S. A. K. Wijethunga†, J. Eldred, C. Y. Tan, E. Pozdeyev, Fermi National Accelerator Laboratory, Batavia, USA

## Abstract

Fermilab Booster synchrotron requires an intensity upgrade from $4.5\times10^{12}$ to $6.5\times10^{12}$ protons per pulse as a part of Fermilab's Proton Improvement Plan-II (PIP-II). One of the factors which may limit the high-intensity performance is the fast transverse instabilities caused by electron cloud effects. According to the experience in the Recycler, the electron cloud gradually builds up over multiple turns in the combined function magnets and can reach final intensities orders of magnitude greater than in a pure dipole. Since the Booster synchrotron also incorporates combined function magnets, it is essential to discover any existence of an electron cloud. And if it does, its effects on the PIP-II era Booster and its mitigating techniques. As the first step, the presence or absence of the electron cloud was investigated using the clearing bunch technique. This paper presents experimental details and observations of the bunch-by-bunch tune shifts of beams with various bunch train structures at low and high intensities and simulation results conducted using PyECLOUD.

## INTRODUCTION

In particle accelerators, free electrons are always present inside the vacuum chambers for many reasons, such as ionization of residual gas molecules, photoemission from the chamber's walls due to synchrotron radiation emitted by the beam, etc. These electrons can be accelerated by the electromagnetic fields of the beam to the energies of several hundreds of eV to a few keV, depending on the beam intensity when such electrons impact vacuum chamber walls, secondary electrons can be generated according to their impact energy and the Secondary Electron Yield (SEY) of the surface. Repeating this process, especially with a proton beam with closely spaced bunches, can lead to an avalanche creating the so-called electron cloud (EC) [1-4].

These ECs can severely limit the performance of high-intensity proton accelerators due to transverse instabilities, transverse emittance growth, particle losses, vacuum degradation, heating of the chamber's surface, etc. The Super Proton Synchrotron (SPS), Proton Synchrotron (PS), and Large Hadron Collider (LHC) at CERN [5-6], Relativistic Heavy Ion Collider (RHIC) at Brookhaven national laboratory (BNL) [7], Proton Storage Ring (PSR) at the Los Alamos National Laboratory (LANL) [8] are few of the high-intensity accelerator facilities that encountered operational challenges due to EC effects.

In 2014, Fermilab Recycler also experienced fast transverse instabilities, and earlier studies by J. Eldred *et al.* [4] indicated that it might be due to the EC build-up in the Recycler. Further investigations by S. A. Antipov [3] showed that the combined function magnets are the EC's most likely buildup locations in the Recycler. The field gradient of the combined function magnets can create a magnetic mirror effect which facilitates electron trapping. According to his simulations, EC accumulates over many revolutions inside a combined function magnet and can reach final intensities orders of magnitudes higher than inside a pure dipole.

The Fermilab Booster [9] is a 474.2 m circumference rapid-cycling (15 Hz) synchrotron containing 96 combined function magnets. It accelerates the beam from 0.4 GeV at injection to 8.0 GeV at extraction over 33.3 ms (the rising portion of the sinusoidal current waveform) in about 20000 turns, where each turn can contain 84 bunches. The proposed Fermilab's Proton Improvement Project-II (PIP-II) requires the Fermilab Booster to deliver a high-intensity beam of $6.5\times10^{12}$ protons per pulse which is a 44% increase in the current intensity [10]. Thus, it is essential to discover any existence of an EC in the PIP-II era Booster, if it does, whether it poses any limitations to the desired performance and its mitigating techniques.

As the first step, the presence or absence of the EC was investigated using the clearing bunch technique. Further, corresponding simulations were carried out with PyECLOUD software [11]. This paper presents the experimental details and observations of the bunch-by-bunch tune shifts of beams with various bunch train structures at low and high intensities and simulation results.

## EXPERIMENTAL TECHNIQUE

According to past observations, a train of closely spaced bunches is required for the electrons to trap in the magnetic field. In the absence of the following bunch, the existing secondary electrons can go through a few elastic reflections and get absorbed by the vacuum chamber. Hence, if a trapped EC is present in the machine, a single bunch following the main batch can be used to clear the EC as it kicks the electrons into the vacuum chamber.

Since an EC act as a lens providing additional focusing or defocusing to the beam, this clearing of the EC can be observed in shifting the betatron tune. According to S. A. Antipov's analysis, a positive tune shift in the horizontal direction indicates the presence of an EC at the beam center, and a negative tune shift in the vertical direction tells the maximum density of the EC near the walls of the vacuum chamber [3]. Adding a clearing bunch can reduce these tune shifts.

To investigate the existence of the EC in the Booster, the clearing bunch technique was used with varying beam intensities, and bunch structures with horizontal and vertical ping and tune shifts were analyzed. This paper presents measurements taken for two different beam intensities (protons per pulse): $4.5\times10^{12}$ (nominal) and $1.9\times10^{12}$ (low). The bunch structure was varied by misaligning the laser notcher and the notcher kicker, as shown in Fig. 1. The Booster employs the laser notcher to remove three bunches out of 84 bunch turn to reduce the losses at the extraction. Since the laser notcher cannot clear the bunches 100%, a kicker pulse is also implemented. In the nominal case, placing the laser notcher and notcher kicker on top of each other resulted in 81 bunch turns. For this study, we misaligned it to have a different bunch structure (opposite notch) with about 79 bunch turns.

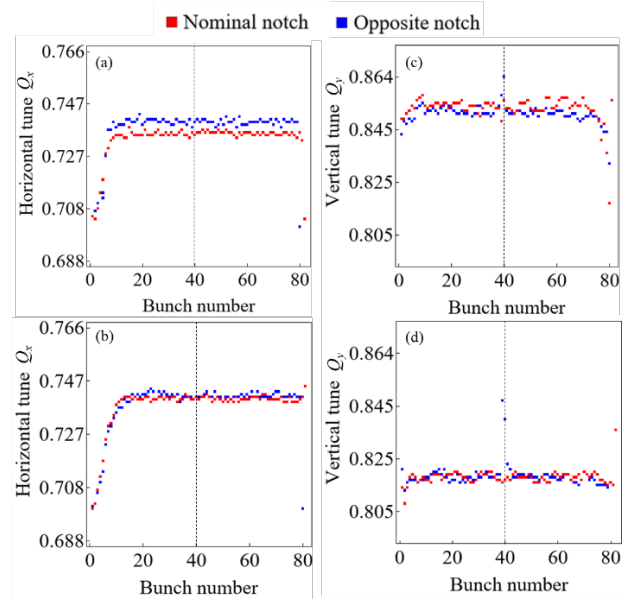

Figure 2: Bunch tune variation in turn 1 for nominal and opposite notches (a) low intensity, horizontal ping, (b) high intensity, horizontal ping, (c) low intensity, vertical ping, and (d) high intensity, vertical ping. The dashed line indicates the location of the notcher.

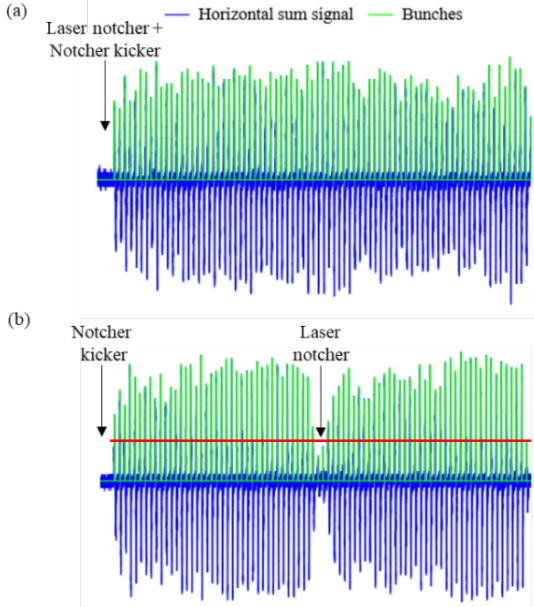

Figure 1: Bunch structures (a) nominal (b) opposite notch. The red line shows the threshold value above which the signal was considered a bunch.

A damper pickup located at long ten was used with a high bandwidth scope to take the measurements. The damper pickup provides the sum and the difference signals for horizontal and vertical directions. The calibration factor is calculated to be 20.0 mm. Before calculating the beam positions, it is essential to ensure each turn of a particular data set follows the same bunch structures corresponding to its notcher pattern. After aligning all the turns, the betatron tune of each bunch of each turn was determined by performing the Fourier transformation.

## RESULTS ANALYSIS

Figure 2 shows each bunch's horizontal and vertical tune variation in turn 1 (near injection) for both bunch structures and both intensities.

The above plots show that the notcher did not affect the horizontal tunes in both high and low-intensity beams. Conversely, the vertical tunes show a positive shift in both high and low-intensity beams due to the notcher. However, this may be due to the impedance tune depression, as the low-intensity beam shows a more minor shift than the high-intensity beam. The more significant horizontal tune shift in the first few bunches cannot be recognized as an effect from the EC as it is too large compared to the Recycler observations and also can be seen in both high and low-intensity beams. According to the observation, this is likely due to the orbital distortion caused by the notcher as it kicks the beam 10 mm in the horizontal direction.

Figure 3 and Fig. 4 depict the horizontal and vertical tune shifts of opposite notch bunch structure concerning the nominal notch bunch structure, from injection up to the transition (~8500 turms), respectively. Note that the tune of each turn was calculated by considering the bunches with tunes unaffected by the notches and taking the average of them. The negative horizontal tune shift near the transition in the high-intensity beam (Fig.3 (a)) reveals that the clearing bunch helped to reduce the tune shift, which is a clear indication of the reduction in EC density. Further, the slight positive vertical tune shift near the transition in the high-intensity beam (Fig.4 (a)) also indicates that the clearing bunch reduces the tune shift, reducing EC density. Low-intensity beam doesn't show a considerable horizontal or vertical tune shift near the transition, which is also consistent with the presence of the EC. The tune shift between 0 to 2000 turns in all the plots could not be fully recognized.

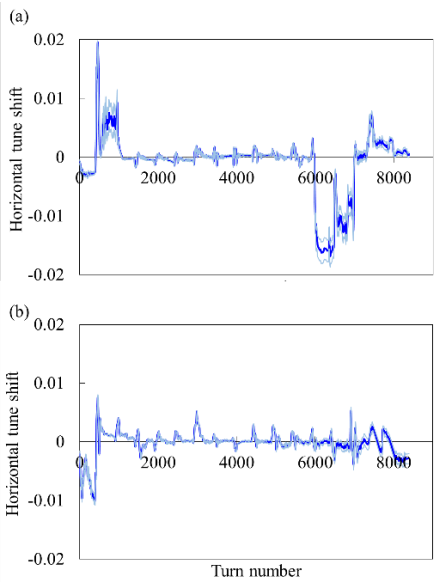

Figure 3: Horizontal tune shift due to the bunch clearing from injection to transition (a) high-intensity, (b) low-intensity. The error was calculated by taking the standard errors of the mean.

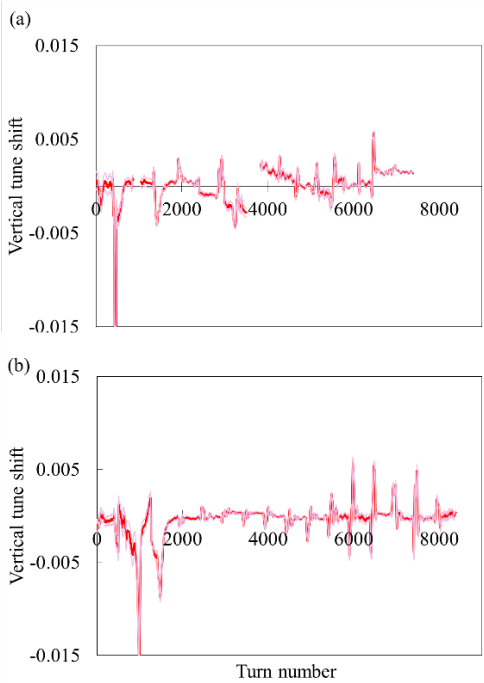

Figure 4: Vertical tune shift due to the bunch clearing from injection to transition (a) high-intensity, (b) low-intensity. The error was calculated by taking the standard errors of the mean. The discontinuity in the high-intensity plot is due to the distorted tune bands.

## SIMULATIONS

To simulate the EC build-up inside a combined function magnetic located in the Booster synchrotron, PyECLOUD code was employed. Table 1 lists the main input parameters used in the simulations.

Table 1: Input parameters in PyECLOUD simulations.

| Parameter | Transition |
|---|---|
| Beam energy [GeV] | 4.2 |
| Bunch spacing [ns] | 19.2 |
| Bunch length, $\sigma$ [m] | 0.253 |
| $\sigma_x, \sigma_y$ [mm] | 3.4, 1.5 |
| SEY, $\delta$ | 1.8 |

The combined function magnet cross-section was considered a rectangle with diploe and quadrupole magnetic fields. The initial number of electrons was taken as $10^4$. The beam filling pattern was included as 81 bunches and 3 empty bunches for the nominal notch and twice 40 bunches and 2 empty bunches for the opposite notch. The simulation was conducted for three turns near transition for both low and high-intensity beams. Figure 5 shows the early simulation results.

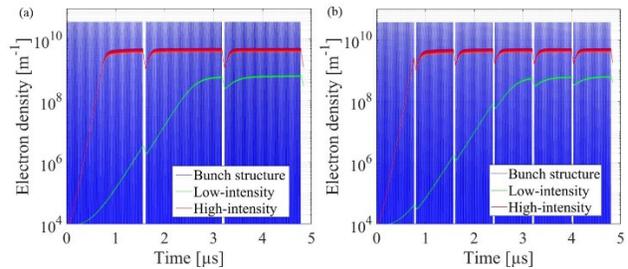

Figure 5: EC build-up for low and high-intensity beams (a) nominal notch, (b) opposite notch.

According to the above plots, EC build-up for low-intensity beams in opposite notch bunch structures is slow compared to the nominal bunch structure. However, both low and high-intensity beam shows almost the same EC saturation despite their bunch structure.

## CONCLUSION

The presence or absence of the EC in the PIP-II era Booster was investigated using the clearing bunch technique. The bunch-by-bunch tune comparison near the injection doesn't show any evidence of EC presence, whereas the average tune comparison near the transition shows some proof of EC presence. PyECLOUD simulations near transition show the EC accumulation doesn't affect by the bunch structure. However, there are a lot of features that we are unable to understand fully. We will continue to further investigate the EC effect in the Booster with more measurements, including microwave measurements and simulations.

## ACKNOWLEDGMENTS

This manuscript has been authored by Fermi Research Alliance, LLC under Contract No. DE-AC02-07CH11359 with the U.S. Department of Energy, Office of Science, Office of High Energy Physics.